\documentclass[acmtog]{acmart}
\renewcommand\footnotetextcopyrightpermission[1]{} 
\usepackage{tikz}
\usepackage{collcell}
\usepackage{pgfplots}
\pgfplotsset{width=8cm,compat=1.9}
\usetikzlibrary{pgfplots.dateplot}
\usepackage{pgfplotstable}
\usepackage{filecontents}
\usepackage{multirow}
\usepackage{caption}
\usepackage{subcaption}
\usepackage{makecell}
    \newcommand*{\MinNumber}{0.0}%
    \newcommand*{\MidNumber}{0.5} %
    \newcommand*{\MaxNumber}{1.0}%
    
    \newcommand{\ApplyGradient}[1]{%
            \ifdim #1 pt > \MidNumber pt
                \pgfmathsetmacro{\PercentColor}{max(min(100.0*(#1 - \MidNumber)/(\MaxNumber-\MidNumber),100.0),0.00)} %
                \hspace{-0.33em}\colorbox{green!\PercentColor!yellow}{#1}
            \else
                \pgfmathsetmacro{\PercentColor}{max(min(100.0*(\MidNumber - #1)/(\MidNumber-\MinNumber),100.0),0.00)} %
                \hspace{-0.33em}\colorbox{red!\PercentColor!yellow}{#1}
            \fi
    }

\pgfplotstableset{
    color cells/.style={
        col sep=comma,
        string type,
        postproc cell content/.code={%
                \pgfkeysalso{@cell content=\rule{0cm}{2.4ex}%
                \pgfmathsetmacro\y{min(100,max(0,abs(round(##1 * 0.5))))}%
                \ifnum##1<0\edef\temp{\noexpand\cellcolor{blue!\y}}\temp\fi%
                \ifnum##1>0\edef\temp{\noexpand\cellcolor{red!\y}}\temp\fi%
                \pgfmathtruncatemacro\x\y%
                \ifnum\x>50 \color{white}\fi%
                ##1}%
                }
    }
}

\AtBeginDocument{%
  \providecommand\BibTeX{{%
    \normalfont B\kern-0.5em{\scshape i\kern-0.25em b}\kern-0.8em\TeX}}}

\setcopyright{none}
\begin{document}


\title{Catching up with trends: The changing landscape of political discussions on twitter in 2014 and 2019}
\author{Avinash Tulasi }
\authornote{These authors contributed equally to work.}
\affiliation{IIIT Delhi}
\email{avinasht@iiitd.ac.in}

\author{Kanay Gupta }
\authornotemark[1]
\affiliation{IIIT Hyderabad}
\email{kanay.gupta@students.iiit.ac.in}

\author{Omkar Gurjar}
\affiliation{IIIT Hyderabad}
\email{omkar.gurjar@students.iiit.ac.in}

\author{Sathvik Sanjeev Buggana}
\affiliation{IIIT Hyderabad}
\email{sathviksanjeev.b@research.iiit.ac.in}

\author{Paras Mehan}
\affiliation{IIIT Delhi}
\email{paras18062@iiitd.ac.in}

\author{Arun Balaji Buduru}
\affiliation{IIIT Delhi}
\email{arunb@iiitd.ac.in}

\author{Ponnurangam Kumaraguru}
\affiliation{IIIT Delhi}
\email{pk@iiitd.ac.in}


\renewcommand{\shortauthors}{Avinash Tulasi and Kanay Gupta, et al.}

\begin{abstract}
The advent of 4G increased the usage of internet in India, which took a huge number of discussions online. Online Social Networks (OSNs) are the center of these discussions. During elections, political discussions constitute a significant portion of the trending topics on these networks. Politicians and political parties catch up with these trends, and social media then becomes a part of their publicity agenda. We cannot ignore this trend in any election, be it the U.S, Germany, France, or India. Twitter is a major platform where we observe these trends. In this work, we examine the magnitude of political discussions on twitter by contrasting the platform usage on levels like gender, political party, and geography, in 2014 and 2019 Indian General Elections. In a further attempt to understand the strategies followed by political parties, we compare twitter usage by Bharatiya Janata Party (BJP) and Indian National Congress (INC) in 2019 General Elections in terms of how efficiently they make use of the platform. We specifically analyze the handles of politicians who emerged victorious. We then proceed to compare political handles held by frontmen of BJP and INC: Narendra Modi (@narendramodi) and Rahul Gandhi (@RahulGandhi) using parameters like "following", "tweeting habits", "sources used to tweet", along with text analysis of tweets. With this work, we also introduce a rich dataset covering a majority of tweets made during the election period in 2014 and 2019. 
\end{abstract}


\maketitle

\section{Introduction}
{

    Internet penetration in India went from 21\% in 2014 to an estimated 34\% in 2019; 420 million users are online via their mobile phones today\cite{SMBP}. The surge in Internet usage resulted in an increase of user base on all popular online platforms. In 2014 the number of users on twitter from India was 15.8 million. This number is at 34.4 million in 2019. While Southern India has the highest amount of social media exposure, this number is comparable to Northern India. The East and West are on the lower side. Growth in usage of the Internet can be attributed to the explosion of 4G cellular networks \cite{doval_2018}. 

    A constant increase in the percentage of people using the Internet has changed how Indians consume news. Interactions and debates have moved from televisions and newspapers to Online Social Networks (OSNs). A variety of research communities from sociology to social network security are showing deep interest in these trends \cite{DBLP:phd/ethos/Fang19}. Changing patterns of user behavior and collective impacts of these changes are being studied in extensive depth \cite{Korakakis2017ASO}. While every aspect of human interaction is essential, elections are of particular interest. The election results impact a nation's future on various fronts like finance, defence, foreign relations, including the day-to-day life of a citizen. India has seen the biggest elections in the world in 2019 with an estimated 900 million eligible voters. A turnout of 67.47\%, when compared to 66.44\% in 2014 General Election \cite{2014_eci_gene} shows a significant increase in participation. Held in seven phases from 11th April 2019 to 23rd May 2019, conducting the election has been one of the most challenging democratic processes in the world. Bharatiya Janata Party (BJP) emerged victorious with Indian National Congress (INC) coming up as the second largest party. Narendra Modi with twitter handle @narendramodi and Rahul Gandhi with twitter handle @RahulGandhi were the leaders of these parties \footnote{We will be using the handles of these leaders to address them in this work}. 
    
    Twitter has had an enormous impact starting from the 2008 Obama campaign and has been an integral part of debates in later U.S. elections that centered around allegations of Russian involvement in 2016 elections\cite{Rezapour2017IdentifyingTO}\cite{Wang2016TacticsAT}\cite{DBLP:conf/comad/DeyKN19}. German \cite{tumasjan2010predicting}, French \cite{sokolova2018elections}, and Italian \cite{vaccari2013social} elections also saw a substantial amount of twitter discussions. 2014 elections in India saw a significant shift to online platforms. There have been studies on 2014 Indian General Elections \cite{ahmed20162014} \cite{khatua2015can} \cite{almatrafi2015application}, showing the effect and size of information flow on twitter \cite{Tumasjan2010PredictingEW}\cite{bhola2014twitter}. With these facts, it is clear that twitter is crucial in an election setting. We proceed to define our Research Questions. 

    \subsection{Research Questions}
    {
         Twitter usage has increased in the years between 2014 and 2019 due to the success of 4G cellular services. Our first question in this research paper is \\
         \textit{\textbf{RQ1}} How has twitter usage in elections changed from 2014 to 2019 ?\\
         
        Tweets, retweets, mentions, and the frequency of tweeting during the election time give rise to establishing a political campaigning strategy. So, our second question is \\
        \textit{\textbf{RQ2} Is there a difference in approach towards campaigning on the twitter platform from 2014 to 2019? }\\

        The style of expression used and focus areas are always different from leader to leader. Impact factor of a tweet depends on how a leader chooses to present the tweet. Usage of simple words and a consistent vocabulary connects to the audience well. Narendra Modi's victory made us curious about the same, so our third question is \\
        \textit{\textbf{RQ3} How do @RahulGandhi's tweets differ from @narendramodi's? }\\
    }

    \subsection{Contributions}
    {
        We performed analysis on twitter data collected from 1st January 2014 to 31st May 2014 and, 1st January 2019 to 31st May 2019, to answer the above research questions. With this work, our contributions are as follows:
        \begin{itemize}
            \item A data set with tweets collected during the 2014 and 2019 General Elections. 
            \item Empirical significance of twitter in Indian General Elections.
            \item Quantitative estimation of the reach of two major parties on twitter. 
            \item Comparison of strategies followed by handles @narendramodi and @RahulGandhi in the 2019 General Election. 
        \end{itemize}
    }

        This paper has five sections. In  Section ~2, we discuss the data used and how we collected this data from the twitter API. Section ~3 compares 2014 and 2019 in terms of behaviour of politicians and winning candidates on twitter,  and how @narendramodi changed his campaigning strategy. In Section ~4, we analyze the general public on twitter, observe the differences between the two major parties, and then compare the reach of @narendramodi and @RahulGandhi based on the content of their tweets. Section ~5 discusses the related work, and in Section ~6, we conclude our work. 

}

\section{Data Description}
{
    We have collected tweets during the election period in 2019 that cover conversations related to elections. In this section, we will describe the methods followed to collect data, and size and variations in the data. Before diving into the collection strategy lets take a look at the Election Schedules of 2014 and 2019 General Elections. 

    \subsection{Election Schedules}
    The 2014 general elections were held in 9 phases from 7th April 2014 to 12th May 2014 ~\cite{2014_eci}. A total of 55,38,01,801 people voted in this election leading to a turnout of 66.40\%. A total of 8,251 candidates contested 543 seats. Results were declared on 16th May 2014, and Narendra Modi became prime minister with Rahul Gandhi as his main opponent \cite{2014_eci_gene}. The 2019 general elections were held in 7 phases from 11th April 2019 to 19th May 2019. 67.11\% turnout was observed in the election, and a total of 8,049 candidates contested 543 \footnote{Election in Vellore constituency was cancelled on 16th April 2019. The by-polls were conducted on 5th August 2019 and results were declared on 9th August 2019.} seats. Results were declared on 23rd May 2019 \cite{2019_eci_gene}.

    \subsection{Data Collection Strategy}
    {
    To make sure we have tweets covering all major topics under discussion, we followed trending hashtag based collection, candidate based collection, and election-day specific collection. Also, we have user snapshots that contain all the tweets on a user's timeline. \\

        \textbf{Hashtag based collection}: Trending hashtags were examined on a daily basis, these hashtags were added to the search pool if they were related to the General Election. Continuous manual supervision in hashtag curating made the process robust and complete. \#LokSabhaElections2019, \#namo are examples of search queries used.  \\
        
        \textbf{Candidate based collection}: Popular political leaders, official handles of political parties and lists of handles submitted by election contestants were used to form a pool of user ids. All tweets by these handles were collected. The collection thus captures entire discussion by the contenders, their sentiments, and any debates that occurred. Using this approach, we collected @narendramodi and @RahulGandhi's tweets. 
        \\
        
        \textbf{Election day tweet collection}: On the seven election days, we collected tweets based on hashtags curated every hour. Frequency of manual hashtag refinement was increased from the earlier strategy. We followed this strategy to capture finer and highly regional sentiments. We were able to capture hashtags like \#OruviralPuratchi, \#isupportgautamgambhir which are significant in Chennai and Delhi. \\

        \textbf{User Snapshots}: For over three months, we have been taking a daily snapshot of user data and tweets by each of these hand-picked users.
    }

    We have used the tweets collected by \cite{bhola2014twitter} for the 2014 analysis part of the paper. In table \ref{stats_main}, we present the size of the data collected. The total number of tweets collected by these methods is 18 million for 2014 and 45 million for 2019 elections. The number of tweets reaching a peak on election days is a typical pattern observed in both these elections.

}
\section{Comparison of twitter usage from 2014 and 2019 in India}
{

\label{sectoin_3.1}

As mentioned in previous sections, the scale of social network usage has increased in India. In this section, we look at how twitter usage has scaled up from 2014 to 2019. 

\subsection{An Overview of political discussions on twitter platform}
{

The number of tweets in our dataset is 18 million in 2014 and 45 million in 2019, as mentioned in table \ref{stats_main}. Average tweets per user are similar for both the years, showing that a larger number of tweets is correlated with the increase in user base.


\begin{center}
    \begin{table}[]
        \centering
        \begin{tabular}{|c|c|c|}
            \hline
                                       & 2014        & 2019        \\
            \hline
            Number of tweets           & 18,705,025 & 45,177,116 \\
            Number of users on twitter & 917,258    & 2,150,179   \\
            Average tweets per user     & 20.39       & 21.01       \\
            \hline
        \end{tabular}
        \caption{Statistics of our dataset from 2014 and 2019, an increase in every aspect of twitter usage is seen here.}
        \label{stats_main}
    \end{table}
\end{center}
}
\subsection{Politicians on twitter}
{
    Table \ref{comp_2014_2019} presents a comparison of twitter usage dynamics in the 2014 and 2019 elections \footnote{Tweets related to the election campaign in 2014 are considered from 1st January 2014 to 31st May 2014. For 2019 this duration is 1st January 2019 to 31st May 2019}. A total of 165 politician handles were captured in our dataset from 2014 \cite{bhola2014twitter}, which was curated manually. For 2019, we extracted twitter handles of contesting candidates from the Election Commission of India (ECI) website, where contesting candidates have provided all their details.  

Out of the 8055 contesting candidates, 1,012 were active on twitter. A 600\% increase in the number of politicians using twitter is observed here. Figure \ref{creation_pol} shows the dates when the twitter handles of politicians were created. Just after the 2014 elections, the profile creation has seen spikes, which indicates the increased awareness among politicians about the twitter platform.

    Twitter provides verified badges to certify the authenticity of a public profile. Many politicians and celebrities chose to get their accounts verified to secure a good reach. Of the 165 political handles that were active on twitter in 2014, 19 handles (11.51\%) were verified. On the other hand, in 2019, the percentage of verified handles is 31.93\%, which is significantly higher.

     Retweets show the reach of a given tweet. While politicians made 23,789 tweets in 2014, the total retweets were 160,469, with an average of 6.74 retweets per tweet.  In 2019, the average number of retweets per tweet was 14.48, which is 114\% higher when compared to 2014. Followers determine the popularity of a person on the platform. Better reach of a politician can thus be correlated with a higher number of followers. In 2014, the total number of followers registered by all politicians was 7,276,249, which makes an average of 44,098.47 per politician. In 2019, an average of 194,298.29 followers per politician is recorded, which was 340\% higher than that of 2014. @narendramodi was the most popular handle in 2014 with 2,432,126 followers. With 47,402,233 followers, @narendramodi is still the most popular handle in 2019. An increase of 20 times in the number of followers is seen here. 

    \begin{figure*}[!t]
        \centering
        \includegraphics{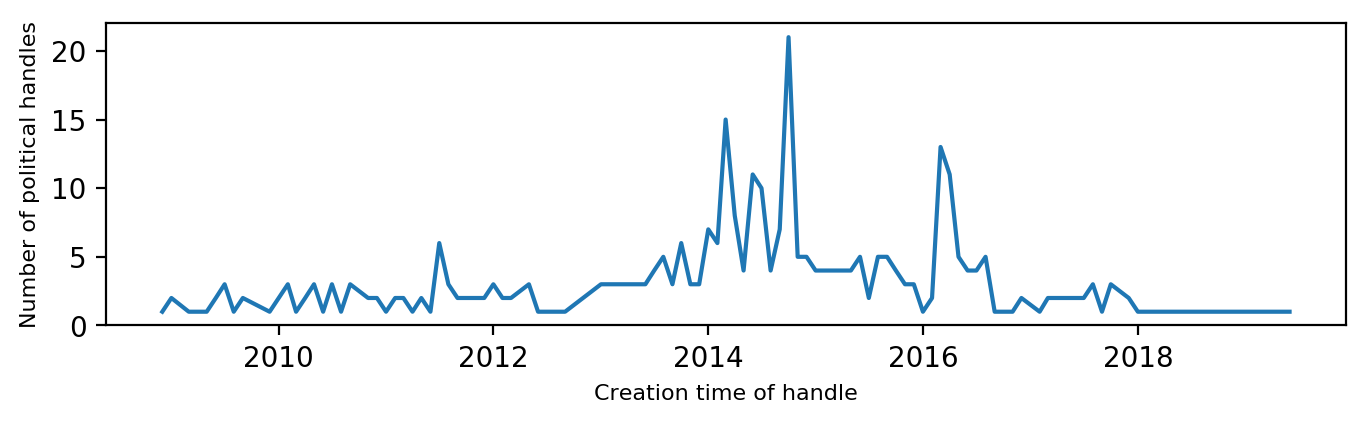}
        \caption{Creation dates of politicians on twitter, spikes after 2014 indicate a raise in inclination towards twitter usage by politicians. }
        \label{creation_pol}
    \end{figure*}

    Just like the public who chose to follow politicians, politicians follow other users on twitter too. On an average, each politician handle followed 116.24 users in 2014. This number increased to 252.50 in 2019. A tweet or retweet made by any user is treated as a status on twitter. 2014 had an average of 144.17 status per politician which decreased to 29.20 in 2019. This indicates that despite the increase in the number of politicians on the platform, the number of tweets did not increase in the same proportion. 

    \textbf{Sources}: In 2014 web and Android were equally used for tweeting, while in 2019 this trend changed and mobile phones became the prime source of tweeting. As seen in figure \ref{source_pol}, this makes up for more than 78.55\% of the total tweet content.

    \textbf{Language}: Every tweet is assigned a language by twitter engine. In our data, the number of tweets made in English for 2014 was 91\% of the total tweets. In 2019 Hindi was the most commonly used language by politicians, which is  53.4\% of the total tweets. English was the next dominant language, with 31.31\% of the total tweets. There was a significant increase in usage of other regional languages like Tamil, Marathi, etc. in 2019 as compared to the 2014 elections. This indicates that politicians started tweeting in local languages to cater to regional audiences.
    
    \begin{table}[!h]
        \centering
        \begin{tabular}{|c | c | c |}
            \hline
                                                          & 2014     & 2019      \\
            \hline
            Number of politicians on twitter              & 165      & 1,012      \\
            Percentage of verified politicians on twitter & 11.51\%  & 31.93\%   \\
            Number of tweets                              & 23,789    & 14,654     \\
            Average number of retweets                    & 6.74     & 14.48     \\
            Average number of followers                   & 44,098.47 & 194,298.19 \\
            Average number of friends                     & 116.24   & 252.50    \\
            Average statuses by politicians               & 144.17   & 29.20     \\
            \hline
        \end{tabular}
        \caption{Comparison of twitter usage and reach of politicians in 2014 and 2019, a general increase in all aspects is seen here.}
        \label{comp_2014_2019}
    \end{table}
    
    \begin{figure}[!h]
        \begin{subfigure}{.25\textwidth}
            \includegraphics[width=\linewidth]{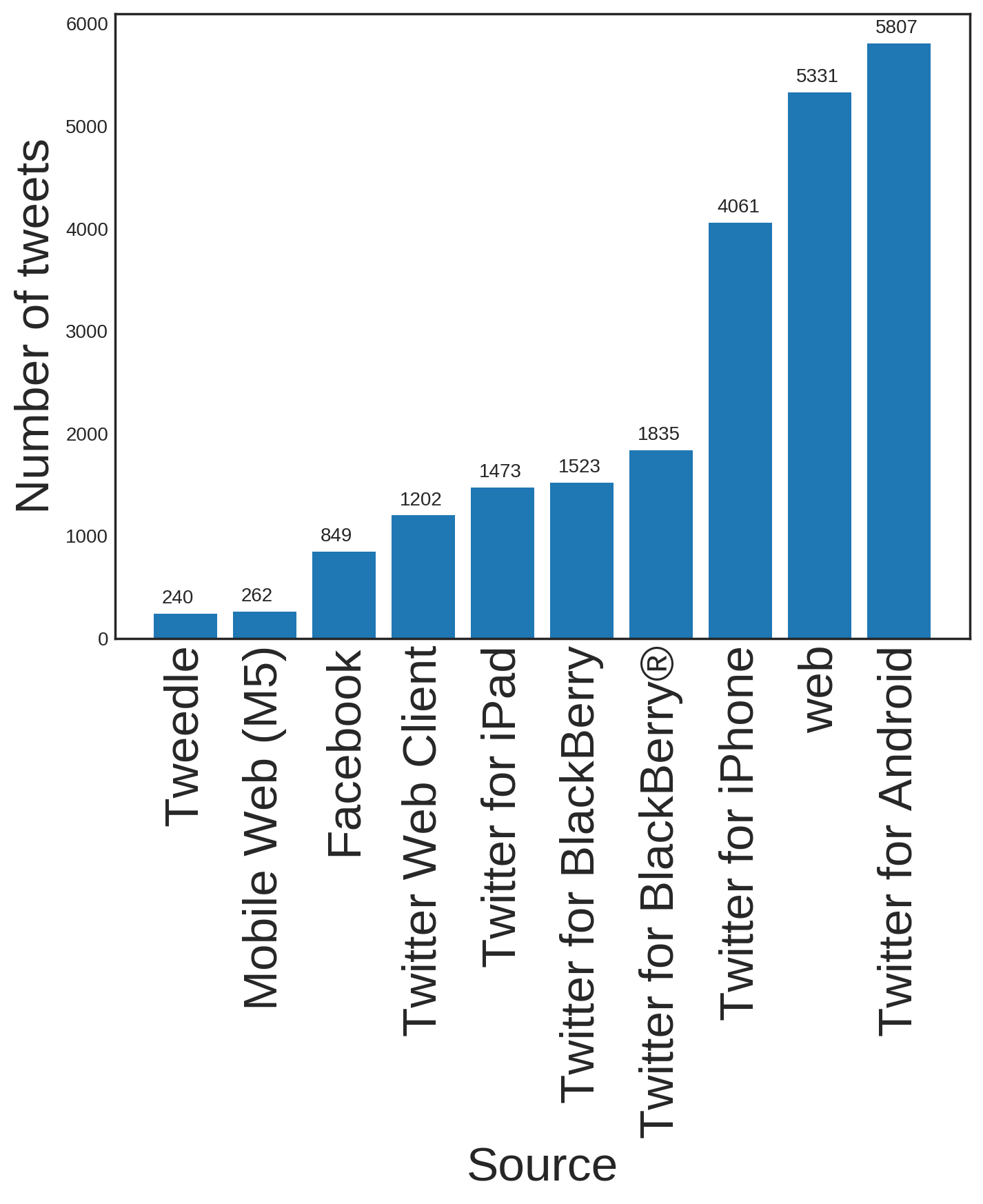}
            \caption{2014}
        \end{subfigure}%
        \begin{subfigure}{.25\textwidth}
            \includegraphics[width=\linewidth]{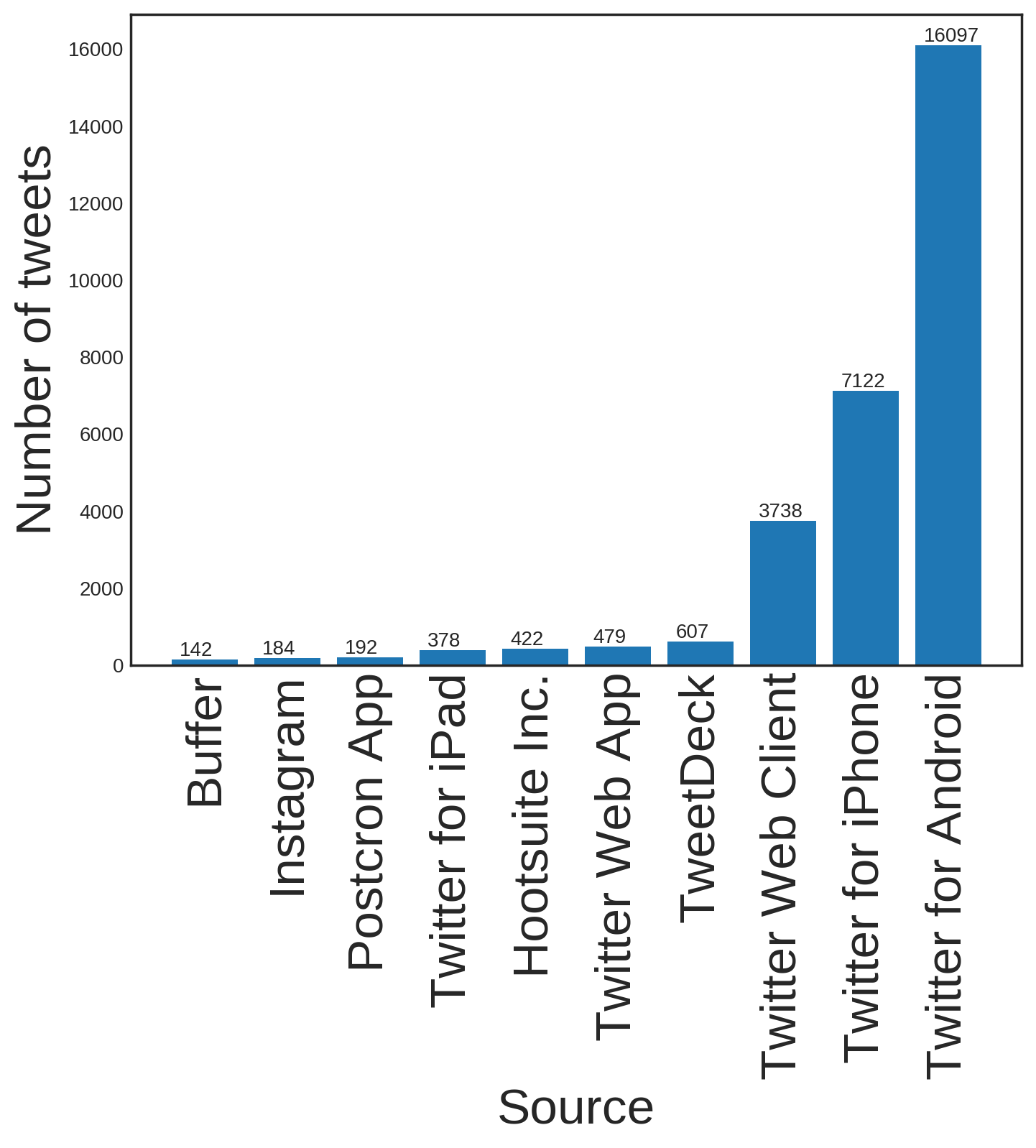}
            \caption{2019}
        \end{subfigure}%
        \caption{Top 10 sources of politician tweets in 2014 and 2019, Android platform is dominant in both 2014 and 2019. Web client usage has seen a sharp decline. }
        \label{source_pol}
    \end{figure}
    
}

\subsection{Winners on twitter}
{
    In this section, we analyze the social media behavior of the 543 winning candidates in 2014 and 2019. We compare the number of candidates active on twitter and the number of winning male and female candidates. Table \ref{gender_var} shows the distribution of male and female winners in 2014 and 2019 General Elections. The number of male candidates contesting in the elections was higher than their female counterparts both in 2014 and 2019, but the difference reduced in 2019 elections. When comparing winning candidates on twitter, the number of male candidates doubled from 2014 to 2019. When comparing the same for verified handles, the number of verified female handles showed a higher increment than that of male handles.\\

    \begin{center}
        \begin{table}[H]
            \centering
            \begin{tabular}{|r|l|r|r|}
                \hline
                                                              &        & 2014 & 2019 \\
                \hline
                \multirow{2}{*}{Number of Winners}            & Male   & 481  & 466  \\
                                                              & Female & 62   & 79   \\
                \hline
                \multirow{2}{*}{Number of Winners on twitter} & Male   & 127  & 281  \\
                                                              & Female & 28   & 49   \\
                \hline
                \multirow{2}{*}{Number of Verified Handles}   & Male   & 91   & 157  \\
                                                              & Female & 14   & 29   \\
                \hline
            \end{tabular}
            \caption{Gender diversity comparison in 2014 and 2019 elections on twitter, a general increase of female winners is seen, but percent females verified are less than males.}
            \label{gender_var}
        \end{table}
    \end{center}
}
\subsection{Comparison of @narendramodi handle in 2014 and 2019}
{
    The table \ref{namo_14_19} presents the statistics of @narendramodi in 2014 and 2019. From the data, the first thing to notice is that the number of tweets has increased by 34.86\%, whereas @narendramodi's statuses increased by 711\% from 2014 to 2019. While the number of users coming online has increased from 2014 to 2019, the followers of the handle @narendramodi also increased proportionally. The number of followers became 200 times more during this period. With the increase in the number of followers the reach of @narendramodi has also increased, this can be derived from the fact that the average number of retweets per tweet went up from a mere 26.63 to a staggering 2985.87.

    However, @narendramodi mentions have seen a substantial fall. While @naremdramodi mentioned 183 unique handles in 2014, this number is just 64 in 2019. These numbers show the increased interaction of @narendramodi on twitter as a mass communication media, as against his earlier method of communication on the platform.

    \begin{table}[]
        \centering
        \begin{tabular}{|c|c|c|}
            \hline
                                        & 2014   & 2019     \\
            \hline
            Tweets between 1-Jan to 31-May & 1,394   & 1,880     \\
            Followers count                & 243,216 & 49,712,260 \\
            Following count                & 795    & 2,227     \\
            Listed count                   & 8,536   & 24,280    \\
            Status count                   & 3,007   & 24,404    \\
            Average retweets count         & 26.63  & 2,985.87  \\
            Unique mentions                & 183    & 64       \\
            \hline
        \end{tabular}
        \caption{Statistics of the handle @narendramodi from 2014 and 2019, tweeting behavior has seen minimal change in numbers, but the increase in followers and statistics related to reach on twitter are clear. }
        \label{namo_14_19}
    \end{table}
    
    @narendramodi's dominant source of tweeting has been the web in both 2014 and 2019. In 2019, there has been a sharp increment in the usage of the source Twitter Media Studio, which shows an increased usage of images and videos in the tweets by this handle.\\
    
    }
}
\section{Twitter in 2019}
{
    In this section, we compare the twitter platform usage by the general public and the two major political parties. First, we look at the geographic distribution of general public, tweet behavior, and language distribution. With these statistics in mind, we discuss the two major political parties, BJP's and INC's footprints on the network. Then we compare the platform usage by the frontmen of BJP (@narendramodi) and INC (@RahulGandhi). Scores like z-score and usage frequency are calculated to make quantitative claims and, the leaders' impact is evaluated based on these scores. 
    
    \subsection{General Public on twitter in 2019}
    {
    
       \begin{figure*}
            \centering
            \includegraphics{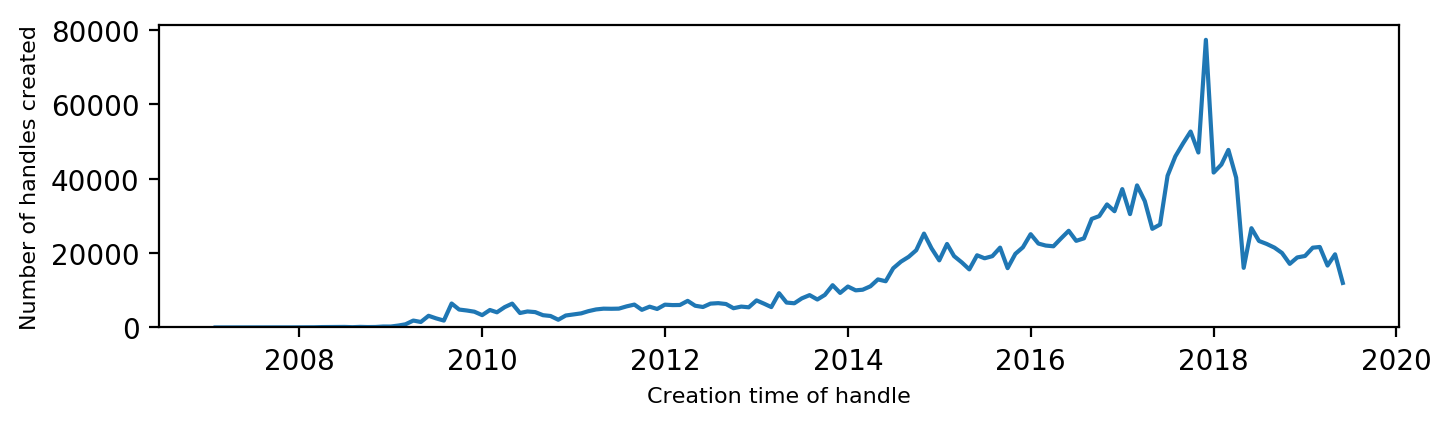}
            \caption{Creation dates of followers of politicians in 2019, in sync with the Internet usage in India, number of profiles created has also increased. }
            \label{followers_pol}
        \end{figure*}
        

        As we have seen in earlier sections, the growth of twitter as a discussion platform has increased steadily, since it was considered to be a platform only for celebrities \cite{twitter_for_celebsonly}.

    Figure \ref{user_stats} shows the creation dates of users in our dataset. It is seen that the platform gained popularity in the late 2017 and early 2018 with a maximum number of profiles created in this period. Given that this data is about the followers of politicians, it can be correlated with the increase in interest in politics.
    In table \ref{user_stats}, we present the statistics of the general public on twitter. A total of 64,283,615 handles were captured in our dataset, with an average of 122.76 tweets per handle. They follow 174.30 twitter handles on an average and are followed by 98.37 handles. Unlike politicians where 31.93\% of the handles are verified, only 0.02\% of the general public handles are verified on twitter. 
        
        \begin{table}[H]
            \centering
            \begin{tabular}{|r|l|}
            \hline
                 Number of users                        & 64,283,615  \\
                 Average number of followers per user   & 98.37     \\
                 Average number of friends per user     & 174.30    \\
                 Average number of statuses per user    & 122.76    \\
                 Percentage verified users              & 0.02 \% \\
            \hline
            \end{tabular}
            \caption{General user statistics on twitter for 2019.}
            \label{user_stats}
        \end{table}
        
    }
    
    \subsection{BJP vs INC}
    {
    
    In this section, we compare the twitter usage of contesting candidates from BJP and INC. Table \ref{tab:my_label} shows the number of candidates who contested from BJP and INC in 2019. While the number of contestants is comparable, both the number of politicians active on twitter and the number of verified handles are higher for BJP. The number of followers for BJP is more than double that of INC.

        \begin{figure}
            \centering
            \includegraphics[scale=0.8]{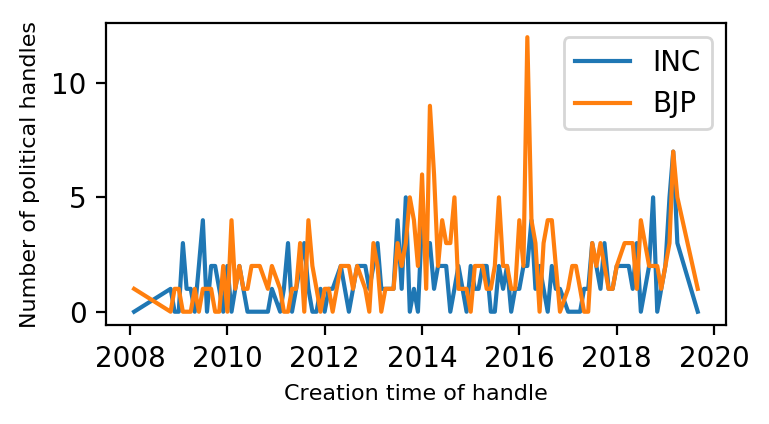}
            \caption{Comparison of profile creation dates of BJP and INC contestants, both parties saw spikes around 2014 and 2016, while number of BJP politicians on twitter are higher. }
            \label{fig:my_label}
        \end{figure}

        \begin{table}[]
            \centering
            \begin{tabular}{|c|c|c|}
            \hline
                   & BJP & INC \\
                   \hline
                  Number of contestants & 436 & 421 \\
                  Number of politicians on twitter & 238 & 167 \\
                  Number of verified handles & 144 & 83 \\
                  Average number of followers & 500,927 & 248,923\\

                  \hline
            \end{tabular}
            \caption{Comparison of BJP and INC contestants on twitter in 2019 elections, with followers number showing a popularity of BJP in public. }
            \label{tab:my_label}
        \end{table}
        
        \begin{figure}
            \centering
            \includegraphics[scale=0.4]{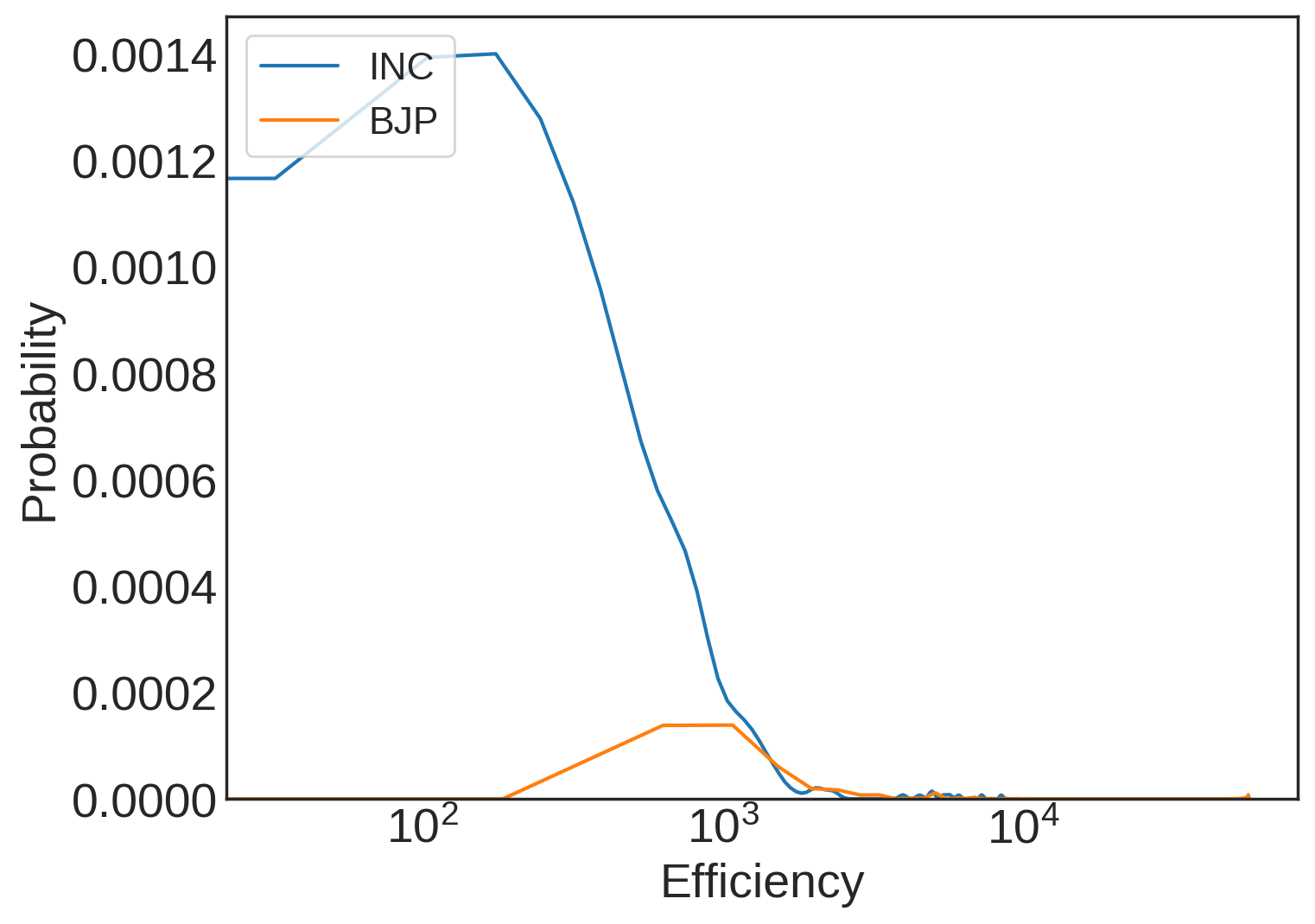}
            \caption{Efficiency comparison of BJP and INC handles, showing BJP handles are more efficient with their reach and platform usage.}
            \label{effi}
        \end{figure}

        \subsubsection{User Mention Analysis}
        {
            Mentions are used to interact with fellow users on the platform. Analyzing mentions can reveal the close ties between different politicians. We use mentions from the tweets of top political handles to generate figure \ref{mentions_20}, which shows the politicians with top 20 followers mentioning each other. To create the graph, we choose edge weight to represent the in-degree, while node size represents the number of mentions a handle made. A bigger size node shows that that handle mentioned a lot of other handles. We can see that the graph is dominant with handles from BJP, with @narendramodi being the most popular with a total of 5,627 mentions. Also, it is the only handle with mentions from all top 20 handles.
        
            \begin{figure}
                \includegraphics[width=\linewidth]{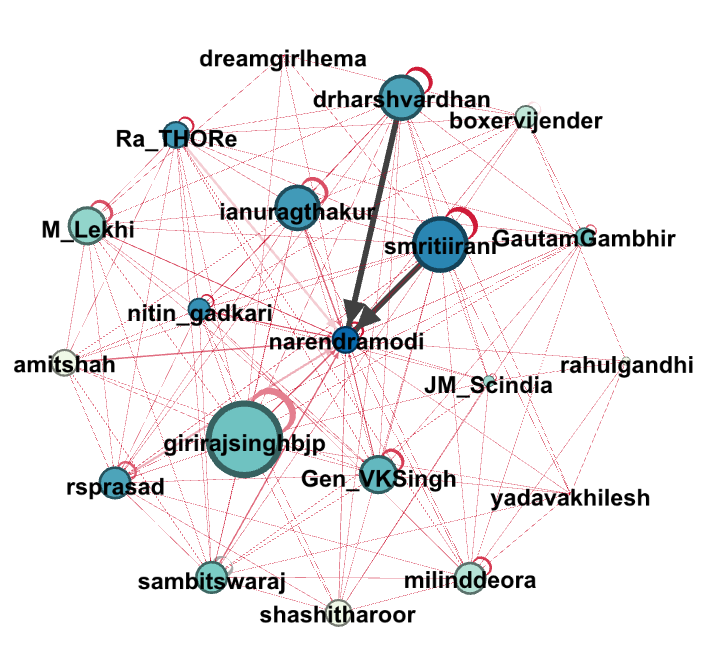}
                \caption{20 Politicians with the most number of followers mentioning each other, @narendramodi being the most mentioned handle, @girirajsingh has mentioned most other handles in top 20, activity of @rahulgandhi is less compared to others in the sphere. }
                \label{mentions_20}
            \end{figure}
        
            While leaders and party are two different entities on twitter, with politicians holding personal handles and parties running their own handles, the public following is more for leaders than parties. This was evident from the observation that amongst the top-20 followed handles, a majority of them belonged to political leaders.
            The mentions by political parties and leaders are interesting to note. @narendramodi mentions the party handle @BJP4India the most. @narendramodi also likes to retweet the tweets that mention itself, and replies to such tweets. Hence, @narendramodi is the second most mentioned handle by @narendramodi. The handle is also frequently mentioned by the top 20. Inspecting the colour scheme, we can see that no other politician has got this number of mentions. @narendramodi is a handle with more popularity than the party @BJP4India. 
        
        }
        
        \subsubsection{Twitter usage efficiency of BJP and INC}
        {
          In this section, we compare the efficiency of politicians from INC and BJP. Efficiency is a measure of how effective users are on a social network. On twitter, users share posts by retweeting them, which essentially adds to the retweet count of the original post. More the count, better is the reach of a post. Based on this fact, we define efficiency as the ratio of the total number of retweets a user gets to the number of tweets made. Mathematically, efficiency can be represented as:
          \[ e_i = \frac{\sum_{j=0}^{n_i}{\#RT_{ij}}}{n_i}\] where $e_i$ is the efficiency of politician $i$ and $\#RT_{ij}$ is the number of retweets user $i$ got on tweet $j$. $n_i$ is the total number of tweets posted by politician $i$. 
        
          A user with efficiency $1$ has as many retweets as tweets. Any number less than  $1$ shows that more than one of the tweets failed to gain any retweets, making them virtually invisible to users on the twitter platform. On the other hand, a number more than $100$ shows visibility of at least a hundred users on the platform. Higher the efficiency score, better is the reach of politicians.
        
          To compare the efficiency of the two contesting parties at a national level, we calculate efficiency of all politicians active at the time of election. On plotting the probability distribution of efficiency in figure \ref{effi}, we can see that while most of the INC politicians are on the near zero side of efficiency, a lot of BJP political handles have a positive efficiency. It is highly probable to find a BJP political handle with a high efficiency value than finding an INC handle. Here, we understand that the BJP handles are more efficient in terms of reach. 
        
        }
    }

    \subsection{@narendramodi vs @RahulGandhi in 2019}
    {
        \begin{center}
        \begin{table}
            \begin{tabular}{|c|c|c|c|}
            \hline
            Handle & Total retweets & Tweet count & Efficiency \\
            \hline
             @narendramodi & 8,549,315 & 1,880 & 4,547.50 \\
            \hline
             @RahulGandhi & 1,619,153 & 221 & 7,326.48\\
            \hline
             
             \hline
            \end{tabular}
            \label{namo_raga}
            \caption{Comparison of number of tweets and retweets of handles @narendramodi and @RahulGandhi.}
        \end{table}
        \end{center}
        
        Given the dominant following and mentions @narendramodi gets, we want to analyze what the tweets are about, and how these tweets different from @RahulGandhi. In this section, we compare @narendramodi and @RahulGandhi using tweet texts on multiple aspects. We study the kind of content posted online and the reach they got. 
        
        \begin{center}
            \begin{table}
                \centering
                \begin{tabular}{|c|c|c|}
                \hline
                                                        & @narendramodi & @RahulGandhi  \\
                \hline
                     Number of followers                & 47,402,233      & 9,731,841       \\
                     Verified followers                 & 8,734          & 2,689          \\
                     \makecell{Followers with \\ at least one tweet}  & 58.46\%      & 63.27\%       \\
                     Average tweets by followers        & 146.85        & 225.48        \\
                     \makecell{Average tweets by followers\\ with at least one tweet}       &  251.17       & 356.36        \\
                               
                     Average followers of followers     & 82.23         & 104.38        \\
                     Average friends of followers       & 107.03        & 156.13        \\
                \hline
                \end{tabular}
                \caption{Follower comparison of @narendramodi and @RahulGandhi. }
                \label{follow_modi_rahul}
            \end{table}
        \end{center}
        
        The number of followers of @narendramodi is 4.87 times more than that of @RahulGandhi. Although this number shows a better reach for the tweets by @narendramodi, the number of users following @narendramodi with at least one recorded tweet is just 58.46\%. However, the number of @RahulGandhi's followers who made at least one tweet are 63.24\%, thus on an average the followers of @RahulGandhi tweeted more. @narendramodi's followers' reach in terms of followers is lesser than that of @RahulGandhi's, but the margin is comparable (82.23 and 104.38). Table \ref{follow_modi_rahul} shows the followers of @narendramodi and @RahulGandhi. While the number of tweets per day is comparable, the retweets @RahulGandhi gets are lesser. A maximum of 243,369 retweets were recorded with Narendramodi's tweets, while the maximum number of retweets is 50,164 for @RahulGandhi. @narendramodi has a better reach on twitter compared to @RahulGandhi. As mentioned in an earlier section, @narendramodi tweets from a web client more than any other source, but @RahulGandhi tweets more from an iPhone.

            \textbf{Linguistic Cues}: 
            Understanding the content and context of tweets requires a close look at the language used by a candidate \cite{labbe2007experiments} \cite{zappavigna2011ambient}. We compare and contrast the language used on twitter by the prime opponents @narendramodi and @RahulGandhi in the 2019 elections. While majority of tweets made by both the handles @narendramodi and @RahulGandhi are in English, the second popular language is Hindi. The proportion of tweets in English to Hindi is more in case of @narendramodi as compared to @RahulGandhi.

    }
    
    \subsubsection{Z-score}
    {
     \begin{center}
        \begin{table}
            \begin{tabular}{ |c|c| } 
            \hline
            @narendramodi & @RahulGandhi \\
            \hline
            government & incindia \\
            development & modiji \\
            india & govt\\
            pmoindia & rahul \\
            narendramodi & gandhi\\
            efforts & congress\\
            nda & rss\\
            people & panjab\\
            shakti & rafale\\
            projects & students\\
            \hline
        \end{tabular}
        \caption{Top words in terms of Z-score, showing the favourite words/topics of @narendramodi and @RahulGandhi.}
        \label{table_zscore}
        \end{table}
        
        \end{center}

        Z-score is used to compare the topic usage patterns \cite{jung2011improving}. The score assumes that the frequency of the text follows a binomial distribution. 
        
        Z-score is defined as :
        \[ Z score (t_{i0}) = \frac{freq(t_{i0}) - n_0 \cdot p(t_i))}{\sqrt{n_0 \cdot p(t_i) \cdot (1 - p(t_i))}}\]
        
        Where $freq(t_{i0})$ corresponds to the frequency of the word $tf_{i}$ in $text-0$ of the first candidate, $n_{0}$ is the number of all the words in the $text-0$, and $pt_i = (tfi0+tfi1) / n.$

        A positive z-score of a word indicates that the word is over-used compared to the other candidate. Similarly, A negative z-score indicates that the word is under-used. 
        \\ \\

        The Table \ref{table_zscore} shows the top +ve z-scores for both the handles. 
        Words like `development,' `programme,' `efforts,' `projects' have the top z-score used by  @narendramodi. Whereas for @RahulGandhi, the top words are different. Words like `rss',`rafale' are on the top with high likelihood. Differences in this table shows the topics of interest and, expected vocabulary used by the opponents.

    }
    \begin{center}
          \begin{table}
          \begin{tabular}{ | c | c | }
          \hline
          Tweet Category & No of tweets \\ 
          \hline
          Campaign / Plan / Manifesto & 797 \\  
          International Events / Talk & 482 \\
          National Interactions & 422 \\
          Project / implementation related & 367 \\
          Congratulatory message to sports / celebrity & 367 \\
          RTs of others & 222\\
          Condolence / Birthday Message  & 190\\
          Opposition Party Talks  & 186\\
          Inspiring Public  & 130\\
          Wishes on festivals & 118\\ 
          Schedule / Plan for travel & 101\\
          PMO Duty & 21\\
          Requesting / asking public to do something & 12\\
          RTs of Media handles & 5\\
          \hline
          \end{tabular}
          \caption{@narendramodi tweets topic Distribution.}
          \label{namo_tweet_content}
          \end{table}
      \end{center}
    
    \begin{center}
          \begin{table}
          \begin{tabular}{ | c | c | }
          \hline
          Tweet Category & No of tweets \\ 
          \hline
          Anti Modi & 231 \\
          Condolence / Birthday Message  & 82\\
          Campaign / Plan / Manifesto & 58 \\  
          Wishes on festivals & 50\\
          Policy / Scheme Opposition & 49 \\
          Schedule / Plan for travel & 39\\
          Congratulating a celebrity & 22 \\
          Inspiring Public  & 22\\
          International Events/Talk & 21 \\
          Requesting / asking public to do something & 15\\
          Cross Party Talks  & 14\\
          RTs of others & 4\\
          Project / implementation related & 3\\
          \hline
          \end{tabular}
          \caption{@RahulGandhi tweets topic Distribution.}
          \label{rahul_tweet_content}
          \end{table}
      \end{center}

   \subsubsection{Tweet content}
    {
       Based on the content of tweets, we classified a sample of tweets into various categories. Tables \ref{namo_tweet_content} and \ref{rahul_tweet_content} show the distribution of tweets across various categories. Categories for both the handles have been kept common except for ones like PMO duty, or opposition party talking about @narendramodi being Anti-Modi. It is interesting to note that Campaign / Plan / Manifesto has the most substantial number of @narendramodi's tweets at about 23\% while @RahulGandhi is seen to have tweeted most Anti-Modi tweets at a 37.2\%. This number is significant in the sense that the second most tweeted category makes for only 13.3\% of the total for @RahulGandhi. While both handles are involved in public relations like congratulating a sports person/celebrity or birthday messages, @narendramodi has a variety of other topics to talk about like international events, national interests like examinations, etc. Also, we notice that the frequency of tweets across these varieties of topics is equal in case of @narendramodi.
@RahulGandhi's tweet content is monotonous as mentioned earlier, making Anti-Modi the most frequent topic to touch on.  \\
        We also looked for categories towards which the users were most receptive using average favourite count. For @narendramodi `Wishes on Festivals' had the greatest average favourite count at about 24,400 favourites per tweet and `RT of others' had the lowest, while in case of @RahulGandhi these were `Inspiring Public' at 33,700 favourites per tweet and `RT of others' respectively. Along with this, we found the classes which induce more discussions using average retweet count as the parameter. The most retweeted category for @narendramodi was `Wishes on festivals' with 4,900 retweets per tweet while for @RahulGandhi it was `Anti-Modi' with a noteworthy average retweet count of 9,000.   
        
    }
}

\section{Related Work}
{
    In this section we present the related work done in elections and twitter domain. Twitter has been an important platform for debate right from Obama's election in 2008 \cite{cogburn2011networked}. Following the interest in this platform, multiple works have discussed the phenomenon in the controversial 2016 U.S. presidential elections \cite{wang2016catching}\cite{Wang2016TacticsAT}. Works studying twitter and elections are not limited to the U.S. political landscape, papers like \cite{Tumasjan2010PredictingEW} studied the German elections, \cite{sokolova2018elections} studied the French elections, \cite{vaccari2013social} studied Italian elections. Indian General Election in 2014 is also studied extensively in \cite{bhola2014twitter} \cite{ahmed20162014} \cite{khatua2015can} \cite{almatrafi2015application}. While the work \cite{kagan2015using} discusses Indian Elections, there is also focus on the Pakistani elections. 
    
    All of the work presented above take information from tweets and analyze different aspects of user/politician behavior in elections, from what users like the most in Trump's tweets \cite{wang2016catching}, to predicting election results based on twitter data in \cite{khatua2015can}. In the work \cite{almatrafi2015application}, the authors analyzed political inclination of users geographically in Indian elections of 2014. While no work related to 2019 Indian General Elections is available yet, multiple blogs and studies like \cite{SMBP} by CSDS, an Indian Government organization, are present. 
}
\section{Discussion}
{
    The scale of twitter usage has increased from 2014 to 2019. This increase is seen in terms of users signing up on twitter and politicians' footprint on twitter. Parameters on the social network platform can be looked at to see the increase in the scale of usage and impact. With an inevitable need to reach the public, politicians have made their presence felt on twitter. Given the increase in scale and usage, we have asked the question if there was any difference in the approach towards campaigning on the twitter platform between 2014 and 2019. In Section \ref{sectoin_3.1}, with the comparison, we saw no major difference in the platform usage, but the campaigning strategy of @narendramodi has seen a drastic change. @narendramodi mentioned some of the most discussed achievements, rather than mentioning fellow party leaders.
    Comparison of the language and linguistic cues answered the third research question. @RahulGandhi's tweets had a variety in terms of topics, vocabulary and sophistication of the language used. However, @narendramodi's simple and consistent language is in stark contrast. As part of the analysis we also compared how @narendramodi gets more mentions than @BJP4India, making @narendramodi an engaging influencer on twitter. 
}
\bibliographystyle{ACM-Reference-Format}
\bibliography{sample-base}

\end{document}